# On The Validity of the Streaming Model for the Redshift-Space Correlation Function in the Linear Regime

Karl B. Fisher

Institute for Advanced Study
School of Natural Sciences
Princeton, NJ 08540

## Abstract

The relation between the galaxy correlation function in real and redshift-space is derived in the linear regime by an appropriate averaging of the joint probability distribution of density and velocity. The derivation recovers the familiar linear theory result on large scales but has the advantage of clearly revealing the dependence of the redshift distortions on the underlying peculiar velocity field; streaming motions give rise to distortions of $\mathcal{O}(\Omega^{0.6}/b)$ while variations in the anisotropic velocity dispersion yield terms of order $\mathcal{O}(\Omega^{1.2}/b^2)$. This probabilistic derivation of the redshift-space correlation function is similar in spirit to the derivation of the commonly used "streaming" model, in which the distortions are given by a convolution of the real-space correlation function with a velocity distribution function. The streaming model is often used to model the redshift-space correlation function on small, highly non-linear, scales. There have been claims in the literature, however, that the streaming model is not valid in the linear regime. Our analysis confirms this claim, but we show that the streaming model can be made consistent with linear theory *provided* that the model for the streaming has the functional form predicted by linear theory *and* that velocity distribution is chosen to be a Gaussian with the correct linear theory dispersion.

# 1 Introduction

Traditionally the two-point correlation function has been the most popular statistic for quantifying galaxy clustering. The correlation function is usually measured from a galaxy redshift catalog by computing pairs in redshift-space. The resulting redshift-space correlation function, $\xi(r_\sigma, r_\pi)$, is a function of pair separations both parallel ($r_\pi$) and perpendicular ($r_\sigma$) to the observer's line of sight; it is an anisotropic function that systematically differs from the real-space correlation function, $\xi(r)$. On small scales, $\xi(r_\sigma, r_\pi)$ is suppressed due to the elongation of non-linear structures along the line of sight, the familiar "Fingers of God." On large scales an opposite effect occurs: coherent in-fall into overdense regions and out-flow from underdense regions compresses structures in redshift-space thereby amplifying $\xi(r_\sigma, r_\pi)$. The redshift-space correlation function thus provides a valuable probe of the peculiar velocity field in both the non-linear and linear regimes. In the gravitational instability picture of structure formation, the amplitude and coherence of the velocity field is directly related to clustering of the underlying mass distribution; by quantifying the distortions in the redshift-space clustering, we can hope to measure important cosmological parameters such as shape of the power spectrum on large scales and the mean mass density, $\Omega$.

There have been two models used in the literature to describe $\xi(r_\sigma, r_\pi)$. The first and perhaps most well known model expresses $\xi(r_\sigma, r_\pi)$ as a convolution of the real-space correlation function with the probability distribution for velocities along the line of sight [1],

$$\xi(r_\sigma, r_\pi) = \int_{-\infty}^{+\infty} dy \, \xi(r) \, F_v\left[r_\pi - y - \frac{y}{r} v_{12}(r)\right] \quad , \tag{1}$$

where $r^2 = y^2 + r_\sigma^2$ and $v_{12}(r)$ is the mean relative peculiar velocity of two galaxies separated by $r$ (cf., Peebles 1980, § 76; Peebles 1993, p. 478). $F_v(V)$ is the probability distribution of relative velocities about $v_{12}(r)$ and is usually assumed to be isotropic with zero mean. By "isotropic" we mean that the tensor second moment of the velocity distribution, $\langle v_i v_j \rangle$, is assumed to of the form, $\sigma^2 \delta_{ij}^K$ ($\delta_{ij}^K$ denotes the usual Kronecker delta symbol). Throughout this Paper, the model for $\xi(r_\sigma, r_\pi)$ given in Equation 1 with an isotropic dispersion will be referred to as the "streaming" model. The streaming model has mainly been used to estimate the galaxy pair-wise velocity dispersion from the suppression of $\xi(r_\sigma, r_\pi)$ on small scales (e.g., Davis & Peebles 1983).

---

[1] We will work in units where the Hubble constant is set to unity, i.e., velocities will have the same units as distances.

The second formulation of the redshift-space correlation function is based on the linear theory of gravitational instability. In linear theory, the peculiar velocity field can be determined directly from the density field; the mapping from real to redshift-space is unique and $\xi(r_\sigma, r_\pi)$ is completely specified. The redshift-space correlation function in linear theory has been calculated by a number of authors (Kaiser 1987; Lilje & Efstathiou 1989; McGill 1990; Hamilton 1992). The result can be compactly written as (Fisher, Scharf, & Lahav 1994) [2]:

$$\xi(r_\sigma, r_\pi) = \left[1 + \frac{2}{3}\beta + \frac{1}{5}\beta^2\right] \xi_0(s) \mathcal{P}_0(\mu) - \left[\frac{4}{3}\beta + \frac{4}{7}\beta^2\right] \xi_2(s) \mathcal{P}_2(\mu) + \left[\frac{8}{35}\beta^2\right] \xi_4(s) \mathcal{P}_4(\mu) \qquad (2)$$

where $\beta = \Omega^{0.6}/b$ is the linear theory growth factor ($b$ is the bias parameter which is assumed to be independent of scale), $s^2 = r_\sigma^2 + r_\pi^2$, and $\mu = r_\pi/s$. The expansion coefficients are given by moments over the power spectrum, $\xi_l(s) = \frac{1}{2\pi^2} \int dk\, k^2 P(k) j_l(ks)$, where $j_l(x)$ and $\mathcal{P}_l(x)$ are the usual spherical Bessel function and Legendre polynomial, respectively, of order $l$. In introducing the bias factor $b$ in Equation 2 we are defining the power spectrum, $P(k)$, to be that of the *galaxy* distribution and not that of the underlying mass, i.e. $P(k) = b^2 P_{mass}(k)$. Consequently, the real-space galaxy correlation function in this notation is simply $\xi(r) = \xi_0(r)$. Throughout this Paper, we will work with quantities of the galaxy distribution; thus $\delta(r)$ will refer to fluctuations in the galaxy distribution and *not* the mass. Consequently factors of the bias parameter, $b$, will arise when we relate velocities (which depend on the mass fluctuations) to $P(k)$. The main interest in modeling $\xi(r_\sigma, r_\pi)$ in the linear regime has been in the possibility of measuring $\beta$ (e.g., Hamilton 1993).

There has been considerable confusion about the compatibility of the two models for $\xi(r_\sigma, r_\pi)$ discussed above. There have been claims in the literature that the streaming model given in Equation 1 fails to asymptote to the linear theory limit represented by Equation 2. Kaiser (1987) and McGill (1990) both pointed out that the streaming model fails to reproduce the terms of $\mathcal{O}(\beta^2)$ in Equation 2. Fisher *et al.* (1994) emphasized that the usual formulation of the streaming model assumes a constant dispersion; they claimed that the missing $\beta^2$ terms could be recovered by generalizing the streaming model to include a scale-dependent dispersion which asymptotically approaches the shape expected in linear theory for large separations.

---

[2] Equation 2 is strictly valid only if the galaxy pairs used to measure $\xi(r_\sigma, r_\pi)$ subtend a small angle with respect to the observer (cf., Cole, Fisher, & Weinberg 1994; Zaroubi & Hofmann 1994); in this Paper we follow Kaiser (1987) and work in this "distant observer" limit.

The purpose of this Paper is to clarify the assumptions that enter the streaming model and show how the model can be made consistent with linear theory on large scales. The streaming model provides an accurate representation of the redshift distortions on small scales, while linear theory is appropriate on large scales. Understanding the relationship between the two approaches will be necessary to construct global models for $\xi(r_\sigma, r_\pi)$ which are applicable on all scales; these models will be needed to if one wishes to take full advantage of the superior data that will be available in the next generation redshift surveys such as the Sloan Digital Sky Survey (Gunn & Knapp 1993; Gunn & Weinberg 1995) and the Anglo-Australian Telescope 2dF galaxy survey.

The outline of this Paper is as follows. In § 2.1, we review the assumptions on which the streaming model is based. A basic inconsistency of the streaming model on large scales is its failure to properly model the density/velocity coupling inherent in linear theory; in § 2.2, we show how this can be rectified by using the the joint density and velocity probability distribution. In § 3, we compute the probability distribution in linear theory and, in § 4 we show that it leads to a redshift-space correlation function which is consistent with the linear theory result given in Equation 2. The derivation reveals the origins of the linear theory distortions in Equation 2: the terms of $\mathcal{O}(\beta)$ are due to scale dependence of the mean streaming along the line of sight while the terms $\mathcal{O}(\beta^2)$ are due to variations in the line of sight velocity dispersion. We conclude in § 5.

## 2 Probabilistic Interpretation of the Redshift Space Correlation Function

### 2.1 Assumptions of the Streaming Model

In order to see clearly the assumptions that enter the streaming model, it is helpful to recall the operational definition of the two-point correlation function in redshift-space, namely that $\xi(r_\sigma, r_\pi)$ represents the excess probability (above Poisson) of finding a galaxy pair (at redshift positions, $\vec{s}$ and $\vec{s}\,'$) with separations $r_\pi$ and $r_\sigma$ parallel and perpendicular to the line of sight,

$$dP = \bar{n}^2\, d^3\vec{s}\, d^3\vec{s}\,' \left[1 + \xi(r_\sigma, r_\pi)\right] \quad , \tag{3}$$

where $\bar{n}$ represents the mean density of galaxies. The streaming model, Equation 1, arises from the conjecture that this probability is equal to the following relationship between the real-space correlation function and velocity distribution function:

$$dP = \bar{n}^2\, d^3\vec{s}\, d^3\vec{s}\,' \left[1 + \xi(r)\right] F_{\rm v}(V)\, \delta^D\left(r_\pi - y - \frac{y}{r} v_{12}(r) - V\right) dV\, dy \quad , \tag{4}$$

where $r^2 = r_\sigma^2 + y^2$. The first term in Equation 4 is the Poisson contribution while $\xi(r)$ represents the excess probability of finding the galaxy pair in real-space due to clustering. The velocity distribution function and Dirac delta function convert the probability of finding separations in real-space to the the desired separations in redshift-space. By equating Equations 3 and 4 and integrating over the unobserved variables $V$ and $y$ one obtains the streaming model given in Equation 1.

There are several points to be made regarding Equation 4:

- The probability of finding the galaxy pair in real-space at separation $r$ is assumed to be independent of the probability of having an associated relative velocity, $V$. This can be seen clearly in Equation 4 where there is an explicit factorization of the the probability of finding the galaxy pair, $\propto [1 + \xi(r)]$, and the velocity distribution function, $F_{\rm v}(V)$. This factorization may be a reasonable approximation on highly non-linear scales where the streaming model is often applied, but as we will see in

§ 2.2, it is not valid in linear theory due to the coupling between the velocity and density fields. The streaming model partially compensates for this failure to model the density/velocity coupling by explicitly imposing a non-zero mean velocity streaming, $v_{12}(r)$, which would vanish in the complete absence of coupling between the density and velocity fields. This ad-hoc introduction of velocity/density coupling makes the streaming model suspect in the linear regime.

- In the usual formulation of the streaming model, the second moment of $F_v(V)$, the velocity dispersion, is taken to be constant (or at most a slowly varying function of separation) and isotropic. However, linear theory predicts that velocities are spatially correlated with a scale-dependent anisotropic dispersion, again casting suspicion of the streaming model on large scales.

- The form of $F_v(V)$ is arbitrary. However, in linear theory the distribution is uniquely specified once the statistical nature of the density field is known. In § 3, we will consider the case where the density field is a Gaussian random field. In this case, the joint distribution of densities and velocities is a multi-variate Gaussian.

## 2.2 Accounting for Density/Velocity Coupling

We now wish to modify Equation 4 to correctly account for any possible velocity/density coupling. This requires knowledge of the joint probability density of measuring both the number densities, $\delta(\vec{r}) = n(\vec{r})/\bar{n} - 1$, and peculiar velocities along the line of sight at two real-space positions, $\vec{x}$ and $\vec{x}\,'$. Let $F_\eta(\vec{\eta})$ denote this joint distribution function with

$$\vec{\eta} = \begin{pmatrix} \delta(\vec{x}) \\ \delta(\vec{x}\,') \\ v_3(\vec{x}) \\ v_3(\vec{x}\,') \end{pmatrix} \quad , \tag{5}$$

and $v_3$ representing the component of the peculiar velocity along the observer's line of sight. In the distant observer approximation, the line of sight can be conveniently taken to lie along the $\vec{x}_3$-axis; the redshift of a galaxy in this limit is just $s(\vec{x}) = \vec{x}_3 + v_3(\vec{x})$.

Since the probability of measuring a galaxy at a position $\vec{r}$ is proportional to the smooth underlying galaxy density, $\propto [1 + \delta(\vec{r})]$, the generalization of Equation 4 can be written down immediately [3]

$$dP = \bar{n}^2 d^3\vec{s}\, d^3\vec{s}\,' \int d^4\vec{\eta}\, dy\, (1+\delta)(1+\delta\,')\, F_\eta(\vec{\eta})\, \delta^D(r_\pi - y - v' + v) \quad , \tag{6}$$

where for convenience we henceforth adopt the abbreviated notation $\delta = \delta(\vec{x})$, $\delta\,' = \delta(\vec{x}\,')$, $v = v_3(\vec{x})$, $v' = v_3(\vec{x}\,')$, $\vec{v} = \vec{v}(\vec{x})$, and $\vec{v}\,' = \vec{v}\,'(\vec{x}\,')$. The integral in Equation 6 averages the joint velocity and density distribution weighted by the probability of actually selecting a galaxy pair with a given redshift separation. Comparing Equations 4 and 6, we obtain the definition of the redshift-space correlation function which correctly accounts for any coupling between the density and velocity fields:

$$1 + \xi(r_\sigma, r_\pi) = \int d^4\vec{\eta}\, dy\, (1+\delta)(1+\delta\,')\, F_\eta(\vec{\eta})\, \delta^D(r_\pi - y - v' + v) \quad . \tag{7}$$

Note that in the special case where the probabilities for the densities and velocities are truly independent, we can write $F_\eta(\vec{\eta}) = P_\delta(\delta, \delta\,') P_v(v, v')$ where $P_\delta$ and $P_v$ are the separate joint distributions for the densities and velocities. In this case, the integral over the densities in Equation 7 just recovers the factor $1 + \xi(r)$ which appears in Equation 4, i.e.,

$$\int d\delta\, d\delta\,' (1+\delta)(1+\delta\,') P_\delta(\delta, \delta\,') = 1 + \xi(r) \quad . \tag{8}$$

## 3 Joint Density/Velocity Distribution in Linear Theory

We now proceed to calculate the joint distribution for the special case where the density field is Gaussian random field and the density and velocity fields are related by linear theory. In this case the joint distribution is a multi-variate Gaussian,

---

[3] A careful reader will note that in addition to the delta function which imposes the correct relative separation, $\vec{s}\,' - \vec{s}$, in redshift-space there should also be a delta function imposing the correct "center of mass," $\vec{s}\,' + \vec{s}$. Translational invariance, however, insures that the probability distribution, $F_\eta(\vec{\eta})$, depends only the redshift separation; the delta function for the center of mass constraint can therefore be immediately integrated out, giving Equation 6.

$$F_\eta(\vec{\eta}) = \frac{1}{(2\pi)^2} \frac{1}{|\det \mathbf{C}|} \exp\left[-\frac{1}{2}\vec{\eta}^\dagger \mathbf{C}^{-1} \vec{\eta}\right] \quad, \tag{9}$$

where $\mathbf{C}$ is the covariance matrix, $\mathbf{C} = \langle \vec{\eta}\vec{\eta}^\dagger \rangle$. The task of computing $F_\eta(\vec{\eta})$ therefore boils down to calculating $\mathbf{C}$ and its inverse; this involves computing the expected coupling between the density and velocity fields.

In linear theory, it is straightforward to show that the density/velocity coupling is given by

$$\langle \delta \, \vec{v}\,' \rangle = -\hat{r} \, \frac{\beta}{2\pi^2} \int dk \, k \, P(k) \, j_1(kr) \quad, \tag{10}$$

where $\vec{r} = \vec{x}\,' - \vec{x}$. Symmetry demands that $\langle \delta \, \vec{v}\,' \rangle$ be directed along the line of separation, $\hat{r}$. Note also that Gaussian fields have the property that the density and velocity *at a point* are uncorrelated, i.e., $\langle \delta \, \vec{v} \rangle = 0$. The bias factor appears in the formula (via $\beta = \Omega^{0.6}/b$) because $\delta$ and $P(k)$ refer to the fluctuations and power spectrum of the galaxy distribution.

It is important to be careful about the meaning of the ensemble expectations values, $\langle \, \rangle$. The Gaussian random fields we are considering are implicitly assumed to be ergodic, and hence ensemble expectation values are equivalent to *spatial* averages. We now wish to compute the average of the galaxy streaming, $v_{12}(r) = \overline{\vec{v}\,' - \vec{v}}$. We must be careful, here, since the average in the case is not a spatial average, but a number-weighted average carried at over *galaxy* positions; since galaxies are located preferentially in high density regions [with an associated probability, $\propto (1 + \delta)$], these two averages need not, and generally are not, the same (cf., Bertschinger 1992). We can recast the number-weighted averages into the familiar spatial averages by explicitly including the probability of finding a galaxy at given position. Thus, we write the expression for the mean streaming as

$$\begin{aligned}
\langle \vec{v}_{12}(\vec{r}) \rangle &\equiv \langle (\vec{v}\,' - \vec{v})(1 + \delta)(1 + \delta\,') \rangle \\
&= \langle \vec{v}\,'\delta \rangle - \langle \vec{v}\delta\,' \rangle + \text{higher order terms} \\
&= -2\,\hat{r} \, \frac{\beta}{2\pi^2} \int dk \, k \, P(k) \, j_1(kr) \quad \text{(in linear theory)} \\
&\equiv v_{12}(r)\,\hat{r} \quad, 
\end{aligned} \tag{11}$$

where the factor $(1 + \delta)(1 + \delta\,')$ represents the effect of number weighting. In Equation 11, the term $\langle \vec{v}\,' - \vec{v} \rangle$ vanishes by symmetry and higher order terms, such as $\langle v \, \delta \, \delta\,' \rangle$, can be dropped in linear theory.

In linear theory there is also a velocity/velocity coupling given by

$$\langle \vec{v}_i \vec{v}_j' \rangle = \Psi_\perp(r) \delta_{ij}^K + \left[ \Psi_\parallel(r) - \Psi_\perp(r) \right] \hat{r}_i \hat{r}_j \quad , \tag{12}$$

where the velocity correlation functions parallel and perpendicular to the line of separation are (Górski 1988)

$$\begin{aligned} \Psi_\perp(r) &= \frac{\beta^2}{2\pi^2} \int dk\, P(k) \frac{j_1(kr)}{kr} \quad , \\ \Psi_\parallel(r) &= \frac{\beta^2}{2\pi^2} \int dk\, P(k) \left[ j_0(kr) - \frac{2 j_1(kr)}{kr} \right] \quad . \end{aligned} \tag{13}$$

Equations 10, 12 along with the definition $\xi(r) = \langle \delta\, \delta' \rangle$ completely specify the elements of the covariance matrix $\mathbf{C}$. At this point we could write down $\mathbf{C}$ and proceed to compute its inverse. Computationally, it is much easier to transform to a new set of variables, $\vec{\eta}'$, in which the covariance matrix is nearly diagonal. Note that the actual probability remains invariant i.e., $F_\eta\, d^4\vec{\eta} = F_{\eta'}\, d^4\vec{\eta}'$. A convenient set of variables are the eigenvectors of the individual density and velocity covariance matrices:

$$\vec{\eta}' = \begin{pmatrix} \delta_+ \\ \delta_- \\ V_+ \\ V_- \end{pmatrix} = \begin{pmatrix} \frac{\delta' + \delta}{\Gamma_+(r)} \\ \frac{\delta' - \delta}{\Gamma_-(r)} \\ \frac{v' + v}{\sigma_+(r)} \\ \frac{v' - v}{\sigma_-(r)} \end{pmatrix} \quad , \tag{14}$$

where

$$\begin{aligned} \Gamma_\pm^2(r) &= 2\left[ \xi(0) \pm \xi(r) \right] \\ \sigma_\pm^2(r) &= 2\left[ \sigma_v^2 \pm \left(\frac{y}{r}\right)^2 \Psi_\parallel(r) \pm \left(\frac{r_\sigma}{r}\right)^2 \Psi_\perp(r) \right] \quad . \end{aligned} \tag{15}$$

In the above equation, $r^2 = y^2 + r_\sigma^2$ with $y$ and $r_\sigma$ being the true separations parallel and perpendicular to the observer's line of sight. $\sigma_v$ is the one dimensional $rms$ velocity,

$$\sigma_v^2 = \frac{1}{3} \cdot \frac{\beta^2}{2\pi^2} \int dk\, P(k) \quad , \tag{16}$$

which by Equation 13 is also given by $\sigma_v^2 = \Psi_\parallel(0) = \Psi_\perp(0)$. The quantity, $\sigma_-(r)$ is merely the pair-wise velocity dispersion along the line of sight.

The covariance matrix expressed in the primed variables is quite simple,

$$\mathbf{C}' = \langle \vec{\eta}\,'\vec{\eta}\,'^{\dagger}\rangle = \begin{pmatrix} 1 & 0 & 0 & \kappa_1 \\ 0 & 1 & -\kappa_2 & 0 \\ 0 & -\kappa_2 & 1 & 0 \\ \kappa_1 & 0 & 0 & 1 \end{pmatrix} \qquad (17)$$

where $\kappa_1 \equiv \frac{y}{r} v_{12}(r)/(\Gamma_+ \sigma_-)$ and $\kappa_2 \equiv \frac{y}{r} v_{12}(r)/(\Gamma_- \sigma_+)$. The advantage of our variable transformation is that it simplifies the resulting expression for the inverse of $\mathbf{C}'$,

$$\mathbf{C}'^{-1} = \begin{pmatrix} \frac{1}{1-\kappa_1^2} & 0 & 0 & \frac{-\kappa_1}{1-\kappa_1^2} \\ 0 & \frac{1}{1-\kappa_2^2} & \frac{\kappa_2}{1-\kappa_2^2} & 0 \\ 0 & \frac{\kappa_2}{1-\kappa_2^2} & \frac{1}{1-\kappa_2^2} & 0 \\ \frac{-\kappa_1}{1-\kappa_1^2} & 0 & 0 & \frac{1}{1-\kappa_1^2} \end{pmatrix}. \qquad (18)$$

Finally, substitution of Equation 18 into Equation 9 gives us the desired joint velocity and density probability distribution in linear theory,

$$F_{\eta'}(\vec{\eta}\,') =$$

$$\frac{1}{(2\pi)^2 (1-\kappa_1^2)^{\frac{1}{2}} (1-\kappa_2^2)^{\frac{1}{2}}} \exp\left[-\frac{1}{2}\left(V_+^2 + V_-^2 + \frac{(\delta_+ - \kappa_1 V_-)^2}{1-\kappa_1^2} + \frac{(\delta_- - \kappa_2 V_+)^2}{1-\kappa_2^2}\right)\right]. \qquad (19)$$

We see that in the limit of weak velocity/density coupling ($\kappa_1$ and $\kappa_2 \ll 1$), the joint probability distribution reduces to two independent Gaussian distributions for the density and velocity fields.

## 4  Linear Theory Limit for $\xi(r_\sigma, r_\pi)$

We can now use the linear theory probability distribution given in Equation 19 to explicitly evaluate the integrals in Equation 7 and thereby obtain the expression for the redshift-space correlation function:

$$1 + \xi(r_\sigma, r_\pi) =$$

$$\int d^4\vec{\eta}\,' dy\, F_{\eta'}(\vec{\eta}\,') \left[1 + \Gamma_+ \delta_+ + \frac{1}{4}\left(\Gamma_+^2 \delta_+^2 - \Gamma_-^2 \delta_-^2\right)\right] \delta^D\left(r_\pi - y - \sigma_+ V_-\right)$$

$$\equiv \int \frac{dy}{\sqrt{2\pi}\sigma_-(r)} G(y) \exp\left[-\frac{1}{2}\frac{(r_\pi - y)^2}{\sigma_-^2(r)}\right], \qquad (20)$$

where once again $r^2 = y^2 + r_\sigma^2$ and the kernel $G(y)$ is defined by

$$G(y) = 1 + \frac{1}{4}\left[\Gamma_+^2 - \Gamma_-^2\right] + \Gamma_+ \kappa_1 \frac{r_\pi - y}{\sigma_-(r)} - \frac{1}{4}\kappa_1^2 \Gamma_+^2 \left[1 - \frac{(r_\pi - y)^2}{\sigma_-^2(r)}\right] \quad . \tag{21}$$

The expression in square brackets in the first line of Equation 20 is just the $(1+\delta)(1+\delta')$ term in Equation 7 expressed in the primed variables. The expression for $G(y)$ is obtained by direct integration over the variables, $d\vec{\eta}' = d\delta_+ \, d\delta_- \, dV_+ \, dV_-$.

To lowest order the exponential and kernel are given by

$$\frac{1}{\sigma_-(r)}\exp\left[-\frac{1}{2}\frac{(r_\pi - y)^2}{\sigma_-^2(r)}\right] \approx \frac{1}{\alpha}\left(1 - \frac{(r_\pi - y)^2}{\alpha^2}\right)\left(\frac{\alpha^2 - \sigma_-^2(y)}{2\alpha^2}\right)\exp\left[-\frac{1}{2}\frac{(r_\pi - y)^2}{\alpha^2}\right] \quad ,$$

$$G(y) \approx 1 + \xi(r) + v_{12}(r)\frac{y}{r}\frac{r_\pi - y}{\alpha^2}, \tag{22}$$

where $\alpha^2 \equiv 2\sigma_v^2$. The redshift-space correlation function to corresponding order is

$$\xi(r_\sigma, r_\pi) =$$

$$\int \frac{dy}{\sqrt{2\pi}\alpha}\left\{\xi(r) + v_{12}(r)\frac{y}{r}\frac{r_\pi - y}{\alpha^2} + \left(1 - \frac{(r_\pi - y)^2}{\alpha^2}\right)\frac{\alpha^2 - \sigma_-^2(y)}{2\alpha^2}\right\}\exp\left[-\frac{1}{2}\frac{(r_\pi - y)^2}{\alpha^2}\right]$$

$$\equiv \int \frac{dy}{\sqrt{2\pi}\alpha}\Xi(y)\exp\left[-\frac{1}{2}\frac{(r_\pi - y)^2}{\alpha^2}\right] \quad , \tag{23}$$

A similar result for the $\xi(r_\sigma, r_\pi)$ in the linear regime has been derived by Regös & Szalay (1995).

To evaluate Equation 23, we note that at separations large compared to the dispersion ($r_\pi \gg \alpha$), the exponential will become sharply peaked at $y = r_\pi$. This suggests that we expand $\Xi(y)$ in a Taylor series about $y = r_\pi$,

$$\begin{aligned}
\xi(r_\sigma, r_\pi) &= \sum_{n=0}^{\infty} \frac{1}{n!}\Xi^{(n)}(r_\pi) \int \frac{dy}{\sqrt{2\pi}\alpha}(y - r_\pi)^n \exp\left[-\frac{1}{2}\frac{(r_\pi - y)^2}{\alpha^2}\right] \\
&= \Xi(r_\pi) + \frac{\alpha^2}{2}\Xi^{(2)}(r_\pi) + \frac{\alpha^4}{8}\Xi^{(4)}(r_\pi) + \text{terms of } \mathcal{O}(\xi^2) \text{ and higher} \\
&= \xi(s) - \frac{d}{dy}\left(v_{12}(r)\frac{y}{r}\right)\bigg|_{y=r_\pi} + \frac{1}{2}\frac{d^2}{dy^2}\left(\sigma_-^2(y)\right)\bigg|_{y=r_\pi} \quad , \tag{24}
\end{aligned}$$

where $\Xi^{(n)}(r_\pi)$ denotes the $n^{\text{th}}$ derivative of $\Xi(y)$ evaluated at $y = r_\pi$. The last line follows from the explicit form of $\Xi(y)$ given in Equation 23 when terms quadratic in the power spectrum, i.e., $\sigma_-^4, v_{12}^2(r)$, etc., are neglected. The last line of Equation 24 shows that the redshift-space correlation function in linear theory differs from the real space correlation function because of variations in both the mean streaming and the pair-wise dispersion along the line of sight, but not by higher order moments of the peculiar velocity field.

Equation 24 is the correct redshift-space correlation function in linear theory. We can cast Equation 24 in the more familiar form of Equation 2 by using the definitions of the streaming and dispersion in terms of the power spectrum (Equations 11 and 15) and then directly computing the derivatives along the line of sight. The calculation is fairly laborious, but the end result is remarkably simple; one finds two equations relating the derivatives of the streaming and dispersion to moments of the power spectrum,

$$
\begin{aligned}
-\frac{d}{dy}\left(\frac{y}{r}v_{12}(r)\right)\Big|_{y=r_\pi} &= \left[\frac{2}{3}\xi_0(s)\mathcal{P}_0(\mu) - \frac{4}{3}\xi_2(s)\mathcal{P}_2(\mu)\right]\beta \\
\frac{1}{2}\frac{d^2}{dy^2}\left(\sigma_-^2(y)\right)\Big|_{y=r_\pi} &= \left[\frac{1}{5}\xi_0(s)\mathcal{P}_0(\mu) - \frac{4}{7}\xi_2(s)\mathcal{P}_2(\mu) + \frac{8}{35}\xi_4(s)\mathcal{P}_4(\mu)\right]\beta^2 \quad . \quad (25)
\end{aligned}
$$

Substitution of the above result into 24 indeed shows that the expressions for $\xi(r_\sigma, r_\pi)$ given in Equations 24 and 2 are equivalent. Equations 2 and 25 are the principal results of this Paper.

## 5 Discussion

In this Paper, we have presented an alternate derivation of the redshift-space correlation function in the linear regime. In contrast with previous derivations based on linear theory, we treat the density and velocity fields in a statistical manner, casting the redshift distortions as appropriate convolutions over the joint density and velocity probability function. The advantage of the probabilistic derivation is that it parallels the formulation of the streaming model which has been frequently employed to model $\xi(r_\sigma, r_\pi)$ on non-linear scales. The derivation shows that the failure of the usual streaming model to reproduce the terms of $\mathcal{O}(\beta^2)$ in linear theory is due to its omission of a scale-dependent dispersion. In linear theory, where we can neglect terms of $\mathcal{O}(\xi^2)$, we can rewrite the correct expression for the redshift-space correlation function (Equation 23) as

$$\xi(r_\sigma, r_\pi) = \int \frac{dy}{\sqrt{2\pi}\sigma_-(r)} \xi(r) \exp\left[-\frac{1}{2}\frac{\left(r_\pi - y - \frac{y}{r}v_{12}(r)\right)^2}{\sigma_-^2(r)}\right] \quad . \tag{26}$$

Comparing this with Equation 4, we see that this is *precisely the streaming model with a Gaussian velocity distribution and a scale-dependent velocity dispersion*. This verifies the claim made by Fisher *et al.* (1994) that the streaming model correctly reproduces linear theory when it generalized to include a scale-dependent velocity dispersion.

The analysis presented suggests a way to modify the streaming model to make it applicable in both the highly non-linear and linear regimes. On highly non-linear scales, one expects the velocity/density coupling will be reduced from its linear theory value, and that the velocity dispersion (e.g., from the virial theorem) will only be a weak function of scale (cf., Peebles 1980, § 76). In such a regime, the standard streaming model with an exponential velocity distribution function provides an excellent fit to the observations (Fisher *et al.* 1994). However, to preserve the correct linear theory behavior at large separations, the distribution function needs to tend towards a Gaussian with mean and dispersion given by the correct linear theory functional forms. One possibility would be to introduce a generalized distribution function with moments that are exponential with isotropic moments on small scales ($\xi(r) \gg 1$) but which tends towards a Gaussian with an anisotropic dispersion in the linear regime ($\xi(r) \ll 1$). It is doubtful if the sampling and depth of current redshift surveys warrant such detailed modeling of the correlation function. With the appearance of much deeper surveys such as the Sloan Digital Sky survey, it will become increasingly important to have models for $\xi(r_\sigma, r_\pi)$ which model both the non-linear and linear regimes.

# Acknowledgments

I would like to thank Michael Strauss, David Weinberg, and David Spergel for stimulating conversations. I gratefully acknowledge the financial support of the Ambrose Monell Foundation.